\documentclass[aps,prl,twocolumn,amsmath,amssymb,nobibnotes]{revtex4-1}%
\usepackage{graphicx}
\usepackage{dcolumn}
\usepackage{bm}
\usepackage{color}
\usepackage{amsmath}
\usepackage{epstopdf}
\usepackage{braket}
\usepackage{fancybox}
\usepackage{amsfonts}
\usepackage{amssymb}%
\usepackage{natbib}
\usepackage[toc,page]{appendix}
\usepackage[symbol]{footmisc}
\renewcommand{\cite}[1]{{[}\onlinecite{#1}{]}}

\newcommand{\be}{\begin{equation}}
\newcommand{\e}{\end{equation}}
\newcommand{\beml}{\begin{subequations}}
\newcommand{\eml}{\end{subequations}}
\newcommand{\beq}{\begin{eqnarray}}
\newcommand{\eq}{\end{eqnarray}}
\newcommand{\ba}{\begin{array}}
\newcommand{\ea}{\end{array}}
\newcommand{\bpm}{\begin{pmatrix}}
\newcommand{\epm}{\end{pmatrix}}
\newcommand{\bc}{\begin{cases}}
\newcommand{\ec}{\end{cases}}

\newcommand{\n}{\nonumber}

\newcommand{\bs}{\boldsymbol}

\begin{document}
\title{Ultrafast control of spin interactions in  honeycomb antiferromagnetic insulators}

\author{Juan M. Losada}
\affiliation{Center for Quantum Spintronics, Department of Physics, Norwegian University of Science and Technology, NO-7491 Trondheim, Norway}
\author{Arne Brataas}
\affiliation{Center for Quantum Spintronics, Department of Physics, Norwegian University of Science and Technology, NO-7491 Trondheim, Norway}
\author{Alireza Qaiumzadeh}
\thanks{Corresponding author: alireza.qaiumzadeh@ntnu.no}
\affiliation{Center for Quantum Spintronics, Department of Physics, Norwegian University of Science and Technology, NO-7491 Trondheim, Norway}

\begin{abstract}
Light enables the ultrafast, direct and nonthermal control of the exchange and Dzyaloshinskii-Moriya interactions. We consider two-dimensional honeycomb lattices described by the Kane-Mele-Hubbard model at half filling and in the strongly correlated regime, i.e., an antiferromagnetic spin-orbit Mott insulator. Based on the Floquet theory, we demonstrate that by changing the amplitude and frequency of polarized laser pulses, one can tune the amplitudes and signs of and even the ratio between the exchange and Dzyaloshinskii-Moriya spin interactions. Furthermore, the renormalizations of the spin interactions are independent of the helicity. Our results pave the way for ultrafast optical spin manipulation in recently discovered two-dimensional magnetic materials.
\end{abstract}

\date{\today}
\maketitle

\textit{Introduction.} The discovery of the all-optical control of the order parameters in antiferromagnetic (AFM) and ferromagnetic (FM) materials by means of ultrashort intense high-frequency laser pulses has propelled spintronics into a new era of ultrafast magnetism \cite{kimel2005ultrafast,2009NatPh...5..727K,Rasing}.
Despite the many attempts to uncover its origin, the detailed underlying microscopic mechanism remains unclear.
The process triggering ultrafast magnetization dynamics phenomena may rely on either thermal or nonthermal mechanisms \cite{heat-light,Rasing}. Thus far, it has been believed that the ultrafast {\it{nonthermal}} manipulation of spins is possible only through either a direct coupling between the magnetic field component of the laser pulses and the spins or an indirect coupling between the electric field component of the light and the spins via spin-orbit coupling \cite{Rasing,Mikhaylovskiy2015,AlirezaQ1,AlirezaQ2}. Recent experiments have demonstrated that laser pulses can directly modify the amplitude and sign of the exchange interaction, which is the strongest spin interaction in magnetically ordered systems \cite{2018Natur.553..481G,Mikhaylovskiy2015,Perakis,PhysRevLett.93.197403}.
References \onlinecite{PhysRevLett.113.057201,PhysRevLett.115.075301,Mentink2015,PhysRevLett.118.097701,Stepanov2017,Kitamura2017,MentinkReview,Barbeau2018} have theoretically proposed that a direct coupling between the electric field of the light and the spins facilitates the nonthermal optical modification of the exchange interaction, in agreement with recent experiments \cite{2018Natur.553..481G,Mikhaylovskiy2015}.

In magnetic systems with broken inversion symmetry, there is also an antisymmetric exchange interaction between spins that breaks the chiral symmetry, namely, the Dzyaloshinskii-Moriya interaction (DMI) \cite{dzyaloshinsky1958, moriya1960,Alireza-DMI1, Alireza-DMI2}. Although this interaction is considerably weaker than the exchange interaction, it is essential in magnetic materials for enabling weak ferromagnetism in AFM materials \cite{dzyaloshinsky1958, moriya1960}, topological objects such as chiral skyrmions \cite{2009Sci...323..915M,2017NatRM...217031F,AlirezaSk1, AlirezaSk2} and chiral domain walls \cite{Thiaville_2012,Parkin,AlirezaSW}, and exotic phases of topological magnon insulators \cite{Owerre2016,PhysRevLett.117.227201, Elyasi2018, Chen2018, Owerre1, Owerre2}. The ratio between the exchange interaction and the DMI controls the tilt angle of the canted spins. Finding a mechanism for tuning this ratio can enable new phenomena in ultrafast spin dynamics and switching \cite{Mikhaylovskiy2015,Mentink2015,Stepanov2017,Kitamura2017}.

Another far-reaching recent breakthrough in spintronics is the discovery of two-dimensional (2D) van der Waals AFM and FM materials with metallic, semiconducting and insulating band structures \cite{2D1,2D2}. The advantages presented by the existence of low dimensionality and magnetic order in the same material enable the development of new spintronic devices with exceptional performance.

In this Rapid Communication, we show that intense high-frequency laser pulses can dramatically affect the spin-spin interactions and the ratio between the exchange interaction and the DMI in a broad class of 2D magnetic materials described by the Kane-Mele-Hubbard model.  We find that light can be used to tune both the sign and magnitude of the AFM exchange interaction, in agreement with the dynamical mean-field theory \cite{Mentink2015}. Importantly, we demonstrate that laser pulses can also be used to independently change the sign and magnitude of the DMI, thus enabling the rapid control of the magnetism. The ability to independently control the signs and magnitudes of the exchange interaction and the DMI enables superior control of magnetic textures in 2D magnets.

\begin{figure}[t]
\centering
\includegraphics[width=0.6\columnwidth]{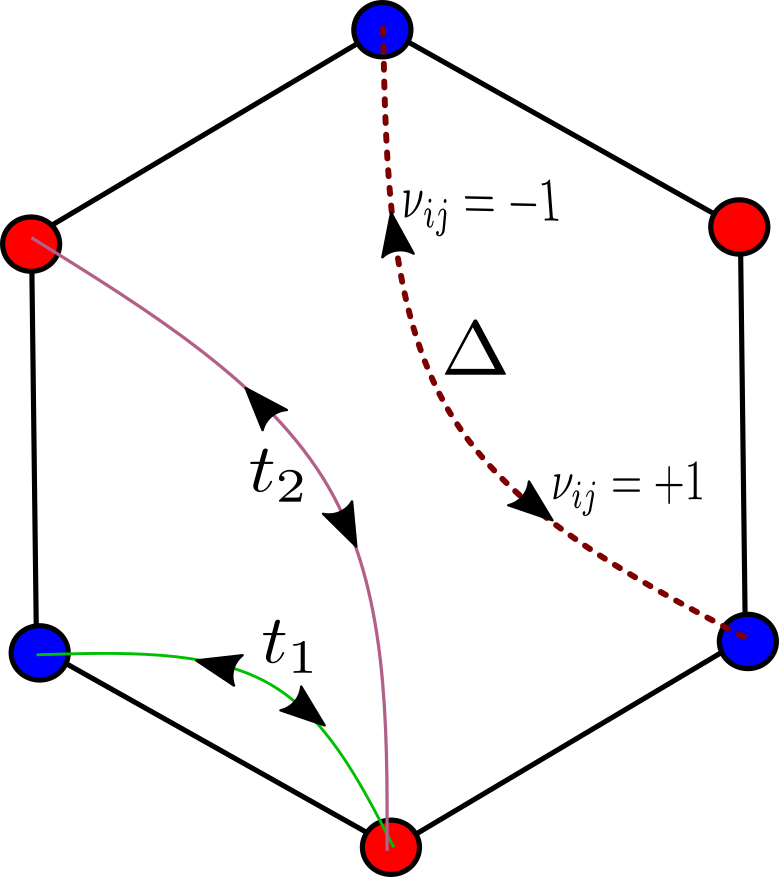}
\caption{A honeycomb cell with NN hopping $t_1$, NNN hopping $t_2$, intrinsic SOI $\Delta$, and $\nu_{ij} = \pm 1$ for clockwise and counterclockwise hopping.}
\label{fig1}
\vspace*{-6pt}
\end{figure}

\textit{Model Hamiltonian.}
The electron dynamics in 2D planar honeycomb lattices can be described by the Kane-Mele-Hubbard model \cite{Kane2005,PhysRevLett.107.010401,PhysRevLett.106.100403,PhysRevB.82.075106,PhysRevB.84.235149,Griset2012,Auerbach}; see Fig. \ref{fig1}. In the absence of external perturbations, the Hamiltonian is the sum of the kinetic term $\hat{\mathcal{H}}_{\text{K}}$, the intrinsic spin-orbit interaction (SOI) $\hat{\mathcal{H}}_{\text{SOI}}$, and the repulsive Coulomb interaction between the electrons as modeled in the form of the extended Hubbard interaction $\hat{\mathcal{H}}_{\text{int}}$,
\begin{equation}
\label{MKMH}
\hat{H}_0 = \hat{\mathcal{H}}_{\text{K}} + \hat{\mathcal{H}}_{\text{SOI}} + \hat{\mathcal{H}}_{\text{int}},
\end{equation}
where
\begin{align}
&\hat{\mathcal{H}}_{\text{K}} = - t_{1}\sum_{\langle i,j \rangle, \tau} \hat{c}_{i \tau}^\dagger \hat{c}_{j \tau} - t_{2}\sum_{\langle \langle i,j \rangle \rangle, \tau} \hat{c}_{i \tau}^\dagger \hat{c}_{j \tau}, \label{Ht} \\
&\hat{\mathcal{H}}_{\text{SOI}}  = i \Delta \sum_{\langle \langle i,j \rangle \rangle, \tau,\tau'} \nu_{ij}\sigma^z_{\tau, \tau'}\hat{c}_{i \tau}^\dagger \hat{c}_{j \tau'} \label{Hsoi}, \\
&\hat{\mathcal{H}}_{\text{int}} = U_{00}\sum_{i=1} \hat{n}_{i\uparrow}\hat{n}_{i\downarrow} + \frac{1}{2}\sum_{\langle i,j \rangle, \tau \tau'} V_{ij}\hat{n}_{i\tau}\hat{n}_{j\tau'} \label{Hint}.
\end{align}
Here, $\langle\cdot\rangle$ and $\langle\langle\cdot\rangle\rangle$ denote nearest neighbors (NN) and next-nearest neighbors (NNN), respectively; $\hat{c}_{i \tau}^\dagger$ and $ \hat{c}_{i \tau}$ are the fermionic creation and annihilation operators, respectively, for an electron at site $i$ and in spin state $\tau=\{\uparrow,\downarrow\}$;
$t_1$ and $t_2$ are the NN and NNN hopping amplitudes, respectively; $\Delta$ is the intrinsic SOI parameter; $\nu_{ij}=\pm 1$, depending on the hopping orientation from $j$ to $i$ (see Fig. \ref{fig1}); $\sigma^{z}$ is the $z$ component of the Pauli matrices $\bm{\sigma}$; and $U_{00}$ and $V_{ij}$ are the on-site and NN Coulomb interactions, respectively.
The intrinsic NNN SOI, Eq. (\ref{Hsoi}), reduces the $\mathrm{SU}(2)$ symmetry of the original Hubbard model to the $\mathrm{U}(1)$ spin group. In buckled noncoplanar honeycomb lattices or systems with structural inversion asymmetry, the presence of NNN or NN Rashba SOIs, respectively, further reduces the symmetry to $Z_2$.

Using the variational principle, it has been shown that the NN Coulomb interaction can be approximated by a renormalized local interaction $U =U_{00} - \bar{V}$, where $\bar{V}$ is a weighted average of the NN Coulomb interaction \cite{Schuler2013}. Thus, we consider only the local Coulomb interaction in our total Hamiltonian and express the interaction part of the Hamiltonian in Eq. (\ref{Hint}) as
\begin{equation}
\hat{\mathcal{H}}_{\text{int}} \approx U\hat{d},
\end{equation}
where we introduce the doublon number operator $\hat{d} = \sum_{i=1} \hat{n}_{i\uparrow}\hat{n}_{i\downarrow}$, which has an eigenvalue of $d$.
For later use, we define $\hat{P}_d$ as the projection operator onto the subspace spanned by states with eigenvalue $d$, i.e., states with exactly $d$ doublons. At site $i$, we can define the projection operator related to double occupancy as $\hat{P}_{i,1} = \hat{n}_{i\uparrow}\hat{n}_{i\downarrow}$ and the projection operator related to the absence of double occupancy as $\hat{P}_{i,0} = 1 - \hat{P}_{i,1}$ \cite{Fazekas}. We can then define the projection operator $\hat{P}_{d}$ for the whole system as follows. Let $\mathcal{O}$ and $\mathcal{P}^d(\mathcal{O})$ denote the set of sites on the lattice and the set of subsets of $\mathcal{O}$ with exactly $d$ elements, respectively. Then, in compact form, the projection operator reads $\hat{P}_d = \sum_{A \in \mathcal{P}^d(\mathcal{O})} \left\{\prod_{i \in A} \hat{P}_{i,1} \prod_{i \notin A} \hat{P}_{i,0} \right\}$.

We are interested in the strongly correlated regime $U\gg t_{1(2)}$ at half filling. In the limit of such strong coupling, any state with a nonzero number of double occupancies ($d \neq 0$) has a much larger energy than a state with no double occupancy ($d=0$). We obtain the effective Hamiltonian acting on the $d=0$ subspace by means of applying second-order perturbation theory on the hopping terms. Using the relations
\begin{subequations}
\label{SpinOperatorInv1}
\begin{align}
&\hat{c}_{i \tau}^\dagger \hat{c}_{i \tau'} = \frac{1}{2} (n_{i \uparrow} + n_{i \downarrow})\delta_{\tau \tau'}  + \bs{S}_i\cdot\bs{\tau}_{\tau', \tau}, \\
&\hat{c}_{i \tau} \hat{c}_{i \tau'}^\dagger = \frac{1}{2} (2 - n_{i \uparrow} - n_{i \downarrow}) \delta_{\tau \tau'} - \bs{S}_i\cdot\bs{\tau}_{\tau, \tau'},
\end{align}
\end{subequations}
we find the spin Hamiltonian for 2D AFM spin-orbit Mott insulators,
\begin{align}
\label{MKMHeff0}
H_{\text{S}} =& J_{1}\sum_{\langle i,j \rangle} \bs{S}_i\cdot\bs{S}_j + J_{2}\sum_{\langle \langle i,j \rangle \rangle} \bs{S}_i\cdot\bs{S}_j,  \n \\
&+ \sum_{\langle \langle i,j \rangle \rangle} \bs{S}_i \bs{\Gamma} \bs{S}_j +\sum_{\langle \langle i,j \rangle \rangle} \bs{D}_{ij}\cdot \bs{S}_i \times \bs{S}_j,
\end{align}
with the following spin-spin interactions,
\begin{subequations}
\label{spin-para}
\begin{align}
&J_{1(2)} = \frac{2t_{1(2)}^2}{U}, \\
&\bs{\Gamma} =\frac{2\Delta^2}{U} \text{diag}(-1,-1,1),\\
&\bs{D}_{ij} = \frac{4 t_2 \Delta}{U}\nu_{ij}  \hat{\mathrm{e}}_z.
\end{align}
\end{subequations}
In the spin Hamiltonian given in Eq. (\ref{MKMHeff0}), the first and second terms are the NN and NNN symmetric Heisenberg AFM exchange interactions ($J_{1(2)}>0$), respectively; the third term is the NNN anisotropic exchange interaction (an XXZ-like term) arising from the intrinsic SOI, and the last term is the intrinsic NNN DMI. The intrinsic SOI in the Kane-Mele-Hubbard model, Eq. (\ref{Hsoi}), leads to an NNN DMI with a DM vector $\bs{D}_{ij}$, which is perpendicular to the honeycomb layer with an amplitude linearly proportional to the SOI strength. It can also be shown that the breaking of the inversion symmetry in this system induces a Rashba-type SOI, which consequently results in an NN interfacial DMI with a DM vector that lies in the film plane and normal to the lattice bonds.

Although the microscopic derivation of the anisotropic exchange interaction in the spin Hamiltonian of Eq. (\ref{MKMHeff0}) has been reported before \cite{PhysRevB.82.075106,Vaezi,2017TDM.....4c5002L}, we are not aware of any other microscopic calculation of the intrinsic NNN DMI in honeycomb lattices \cite{Owerre2016,PhysRevLett.117.227201,Chen2018}. The spin Hamiltonian of Eq. (\ref{MKMHeff0}) gives rise to several interesting features and exotic phases, such as the existence of the magnon spin Nernst effect in collinear AFM layers \cite{Cheng2016, PhysRevLett.117.217203}, a topological magnon insulator phase \cite{Owerre2016, PhysRevLett.117.227201, Elyasi2018, Chen2018}, spin Hall effects for Weyl magnons \cite{Zyuzin2018, Sekine2016}, magnonic Floquet topological insulators, spin density waves \cite{Mulder2010}, and chiral and topological gapped spin liquid phases \cite{Vaezi}. The ultrafast control of the DMI and the exchange interaction, by means of laser pulses, can enable the engineering of all of these phenomena and phases.

For completeness, let us briefly illustrate the effect of disorder by adding an onsite disorder potential $\sum_{i \tau} \varepsilon_i \hat{c}_{i \tau}^\dagger \hat{c}_{i \tau}$ to the Kane-Mele-Hubbard Hamiltonian given in Eq. (\ref{MKMH}), where $\varepsilon_i$ is an uncorrelated random variable. Following the above procedure, it can be shown that the spin interaction parameters are renormalized as $1/U \rightarrow 1/[U-(\epsilon_j-\epsilon_i)^2/U]$ \cite{Protopopov2018}. In the large $U$ limit and in the presence of very high frequency oscillations, the effect of disorder is negligible; thus, we do not include it in the following.

\textit{Laser illumination.}
We introduce the effects of laser irradiation in the Kane-Mele-Hubbard Hamiltonian of Eq. (\ref{MKMH}) via the Peierls substitution \cite{Peierls1933}. The Peierls' prescription is valid for slowly varying vector potentials on the scale of the lattice constant at which the system remains in a quasiequilibrium state \cite{PhysRev.84.814}. The electric field component of a polarized laser pulse is given by  $\bs{E}(t) = E_0( e^{-i\omega t}\hat{\epsilon}+\mathrm{c.c.})/2$, where $E_0$ is the electric field amplitude; $\omega$ is the laser pulse frequency; and $\hat{\epsilon} = (\hat{\mathrm{e}}_x+i\lambda\hat{\mathrm{e}}_y)/\sqrt{1+\lambda^2}$ is the unit vector representing the laser polarization, with $\lambda= 0$ for linear polarization and $\lambda= \pm 1$ for right- and left-handed polarizations.

It is convenient to rewrite the noninteracting part of the Kane-Mele-Hubbard Hamiltonian in Eq. (\ref{MKMH}) as an effective hopping term $\hat{T}_0=\hat{\mathcal{H}}_{\text{K}}+\hat{\mathcal{H}}_{\text{SOI}} = - \sum_{i,j , \tau, \tau'}
t_{ij}^{\tau\tau'} \hat{c}_{i \tau}^\dagger \hat{c}_{j \tau'}$, where the hopping amplitude is $t_{ij}^{\tau\tau'} = \delta_{\tau,\tau'}t_1$ for $i$ and $j$ that satisfy the NN condition and $t_{ij}^{\tau\tau'} = \delta_{\tau,\tau'}t_2 - i\Delta\nu_{ij}\sigma^z_{\tau, \tau'}$ for $i$ and $j$ that satisfy the NNN condition. With the Peierls substitution, the hopping part of the Hamiltonian gains an extra phase $t_{ij}^{\tau\tau'}\rightarrow t_{ij}^{\tau\tau'} e^{{i \frac{e}{\hbar} \bs{R}_{ij} \cdot \bs{A}(t)}}$, where $\bs{R}_{ij} = \bs{R}_i-\bs{R}_j$, $\bs{R}_i$ is the position of site $i$, $e$ is the charge of an electron, $\hbar$ is the reduced Planck constant and $\bs{A}$ is the vector potential of the laser pulse; $\bs{A}(t) = \frac{1}{2}(\bs{A} e^{-i\omega t} + \mathrm{c.c.})$, with $\bs{A} = \frac{iE_0}{\omega}\hat{\epsilon}$.
The Peierls phase at time $t=0$ can be rewritten as $\frac{e}{\hbar}\bs{R}_{ij}\cdot\bs{A} \equiv \alpha_{ij} e^{i \theta_{ij}}$, with $\alpha_{ij} = \pm|\frac{e}{\hbar}\bs{R}_{ij}\cdot \bs{A}|$, such that $\alpha_{ij}= -\alpha_{ji}$, $\theta_{ij}= \theta_{ji}$, and $\theta_{ij} \in \left[0,\pi\right)$.
Now, we can use the Jacobi-Anger expansion to rewrite the Peierls phase in the basis of its harmonics:
\begin{equation}
\label{JacobiAnger}
e^{i\frac{e}{\hbar}\bs{R}_{ij}\cdot\bs{A}(t)} = \sum_m e^{i(\frac{\pi}{2}-\theta_{ij})m} \mathcal{J}_m(\alpha_{ij}) e^{im\omega t},
\end{equation}
where $\mathcal{J}_m(x)$ is an \textit{m}th Bessel function of the first kind \cite{Kitamura2017}.

In the presence of the laser field, the hopping term in the Hamiltonian depends on time, $\hat{H}(t) = \hat{T}(t) +  U\hat{d}$. From Eq. (\ref{JacobiAnger}), we find that $\hat{T}(t) = \sum_m \hat{T}_m e^{im \omega t}$, where $\hat{T}_m$ is the sum of all \textit{m}-th Fourier modes of the hopping terms. We can additionally adopt the decomposition $\hat{T}_m = \hat{T}_{-1,m}+\hat{T}_{0,m}+\hat{T}_{1,m}$, where $\hat{T}_{dm}(t)$ changes the doublon number by adding $d$ double occupancies and is expressed as $\hat{T}_{dm}(t) = \sum_n \hat{P}_{n+d}\hat{T}_{m}(t)\hat{P}_n$. Since the hopping term is of second order in the creation and annihilation operators, it can change the double occupancy of the states only by $\pm1$. Thus, we can express the hopping operator as
\begin{equation}
\hat{T}(t) = \sum_m (\hat{T}_{-1,m}+\hat{T}_{0,m}+\hat{T}_{1,m})e^{im\omega t}.
\end{equation}

To find the renormalized spin Hamiltonian in the strongly correlated regime, we first derive an effective static Hamiltonian using the Floquet formalism \cite{Floquet1,2018NJPh...20i3022R,Floquet2}. To this end, we transform the original time-dependent Hamiltonian $\hat{H}(t)$ by using the canonical transformation $\hat{U}(t) = e^{-i\hat{S}(t)}$ \cite{PhysRevLett.116.125301,Kitamura2017},
\begin{equation}
\hat{H}'(t) = e^{i\hat{S}(t)} [\hat{H}(t)  -  i\partial_t] e^{-i\hat{S}(t)}.
\label{Htransformed}
\end{equation}
We can formally express $\hat{T}(t) = \eta \hat{T}(t)$, where $\eta$ plays the role of a bookkeeping parameter in the perturbation expansion. We expand $\hat{S}(t) = \sum_\nu \eta^\nu \hat{S}^{(\nu)}(t)$ and $\hat{H}'(t) = \sum_\nu \eta^\nu \hat{H}'^{(\nu)}(t)$. We require the transformed Hamiltonian to be block diagonal in the doublon number operator $\hat{d}$. To fulfill this requirement, the unitary transformation $\hat{S}(t)$ must have the same periodicity as $\hat{T}(t)$; consequently, the transformed Hamiltonian $\hat{H}'(t)$ will have the same periodicity as the original Hamiltonian $\hat{H}(t)$. Thus, we can write $\hat{S}^{(\nu)}(t) = \sum_m e^{im\omega t}\hat{S}^{(\nu)}_m$. With the further requirement that $\hat{S}(t)$ does not contain block-diagonal terms, we can uniquely determine the unitary transformation,
\begin{equation}
\hat{S}^{(\nu)}(t) = \sum_{d \neq 0} \sum_m \eta^\nu \hat{S}^{(\nu)}_{d,m} e^{im\omega t},
\end{equation}
where $\hat{S}^{(\nu)}_{d,m}$ changes the double occupancy number by $d$. We expand the transformed Hamiltonian of Eq. (\ref{Htransformed}) into a power series in $\eta$ and determine $\hat{S}^{(\nu)}(t)$ iteratively in $\nu$ such that $\hat{H}'^{(\nu)}(t)$ is diagonal in the doublon number. After tedious but straightforward calculations, we obtain the transformed Hamiltonian up to the second order in the hopping parameter, $\hat{H}'(t)= \hat{T}'(t)+U\hat{d}$, where
\begin{align}
\label{transformedH}
&\hat{T}'(t) \approx  - \sum_m \hat{T}_{0,m}(t)e^{im\omega t}  \n \\
&+ \frac{1}{2}\sum_{mn} \left( \frac{\left[\hat{T}_{1,n}, \hat{T}_{-1,m-n} \right]}{U+n\hbar\omega} - \frac{\left[\hat{T}_{-1,n}, \hat{T}_{1,m-n} \right]}{U-n\hbar\omega} \right) e^{im\omega t}.
\end{align}
Now, we calculate the effective static Hamiltonian by time averaging the transformed Hamiltonian $\hat{H}_{\text{eff}}=\hat{P}_0\hat{H}'(t)\hat{P}_0$, where $\hat{P}_0$ is the Gutzwiller projection onto the subspace containing no doubly occupied sites at all, i.e., the $d=0$ subspace \cite{Fazekas}.
After some algebra, the effective static Hamiltonian is obtained in terms of the creation and annihilation operators:
\begin{align}
\hat{H}_{\text{eff}} &= - \sum_{i,j, \tau, \tau'} \left(t_{ij}^{\tau} t_{ji}^{\tau'} \sum_{n} \frac{\mathcal{J}_{n}^2(\alpha_{ij})}{U+n\hbar\omega} \right)  \hat{c}_{i \tau}^\dagger \hat{c}_{j \tau} \hat{c}_{j \tau'}^\dagger \hat{c}_{i \tau'}. \label{GeneralHeff}
\end{align}

Note that to obtain this result, we have only considered the strongly correlated regime of the Kane-Mele-Hubbard model, $t_{1(2)}/U \ll 1$, and no assumption has been made on the range of $\omega$ and $\alpha_{ij}$. Using the relations in Eq. (\ref{SpinOperatorInv1}), we finally obtain the spin Hamiltonian at half filling,
\begin{align}
\label{MKMHeffw}
\tilde{H}_{\text{S}}(\omega) =& \sum_{\langle i,j \rangle} \tilde{J}_{1,ij} \bs{S}_i\cdot\bs{S}_j + \sum_{\langle \langle i,j \rangle \rangle} \tilde{J}_{2,ij}\bs{S}_i\cdot\bs{S}_j \n \\
&+ \sum_{\langle \langle i,j \rangle \rangle} \bs{S}_i \tilde{\bs{\Gamma}}_{ij} \bs{S}_j +\sum_{\langle \langle i,j \rangle \rangle} \tilde{\bs{D}}_{ij}\cdot \bs{S}_i \times \bs{S}_j,
\end{align}
with the following renormalized spin-spin interactions,
\begin{subequations}
\label{Renor-para}
\begin{align}
&\tilde{J}_{1, ij} = 2t_{1}^2\sum_n\frac{\mathcal{J}_{n}^2(\alpha_{\langle ij\rangle})}{U+n\hbar\omega}, \\
&\tilde{J}_{2, ij } = 2t_{2}^2\sum_n\frac{\mathcal{J}_{n}^2(\alpha_{\langle\langle ij\rangle\rangle})}{U+n\hbar\omega}, \\
&\tilde{\bs{\Gamma}}_{ ij} = 2\Delta^2 \text{diag}(-1,-1,1) \sum_n\frac{\mathcal{J}_{n}^2(\alpha_{\langle\langle ij\rangle\rangle})}{U+n\hbar\omega},\\
&\tilde{\bs{D}}_{ ij} = 4 t_2 \Delta \sum_n\frac{\mathcal{J}_{n}^2(\alpha_{\langle\langle ij\rangle\rangle})}{U+n\hbar\omega} \nu_{ij} \hat{\mathrm{e}}_z.
\end{align}
\end{subequations}
All spin interaction parameters are renormalized in the presence of a periodic drive by the same function, but $\alpha_{ij}$ differs between the NN and NNN parameters. Thus, the ratios between the renormalized NN and NNN parameters are different from those for the unperturbed parameters. Therefore, in a honeycomb lattice described by the Kane-Mele-Hubbard model, the ratio between the AFM exchange interaction and the intrinsic DMI changes during the light irradiation, while in a square lattice with the NN Rashba SOI, it is not possible to control this ratio.
The renormalized spin interactions presented in Eq. (\ref{Renor-para}) do not depend on the helicity in this model. In our model, perturbing the system with an AC electric field renormalizes the original DMI, which is already present in the unperturbed Hamiltonian. In the absence of a SOI, the unperturbed system does not display DMI and adding the electric field would not induce any DMI. This is different from the case studied in Ref. \cite{PhysRevLett.116.125301}, in which it has been shown that an out-of-plane AC electric field, equivalent to a periodic time-dependent chemical potential, induces a DMI-like term in the system even in the absence of any SOI.

\begin{figure}[t]
\centering
\vspace{-1.26cm}
\includegraphics[width=\columnwidth]{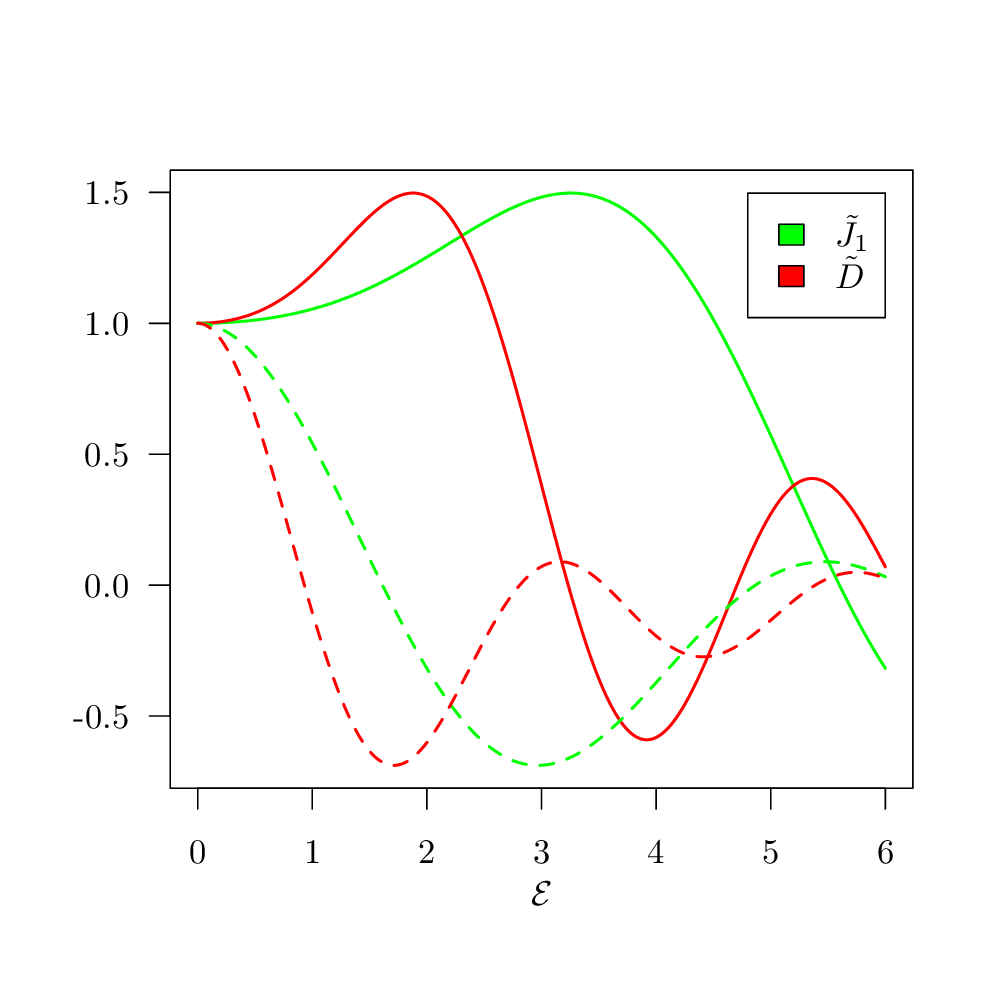}
\vspace{-1.26cm}
\caption{Dimensionless exchange interaction (green lines) and DMI (red lines) as functions of the Floquet parameter  $\mathcal{E} = \frac{eaE_0}{\hbar \omega}$ for two laser frequencies, $\omega = 4$ (solid lines) and $\omega = 14$ (dashed lines).}
\label{fig2}
\end{figure}

Figure \ref{fig2} shows the dependence of the dimensionless NN exchange interaction $\tilde{J}_{1,ij}/J_{1,ij}$ and the dimensionless NNN DMI $\tilde{D}_{ij}/D_{ij}$ on the Floquet parameter $\mathcal{E} = \frac{e a E_0}{\hbar\omega}$, where $a$ is the lattice constant. We show this dependence for two different laser pulse frequencies. We set $\hbar=t_1 = 1$ and measure the energy in units of $t_1$ and the frequency in units of $t_1/\hbar$. The presented results correspond to the material parameters $t_2=0.1$ and $U = 10$. Figure \ref{fig2} shows that it is possible to not only change the sign and amplitude of the exchange interaction, as reported in Ref. \cite{Mentink2015}, but also change the sign and amplitude of the intrinsic DMI. Figure \ref{fig2} shows that the ratio $\tilde{D}_{ij}/\tilde{J}_{1,ij}\neq D_{ij}/J_{1,ij}$, which is responsible for ultrafast photoinduced spin dynamics phenomena, can be tuned by means of laser excitations in systems with specific symmetries. Here, we should emphasize that since in our model the NNN exchange interaction, the anisotropic exchange interaction, and the intrinsic DMI arise from NNN couplings, they are renormalized in the same way [see Eqs. \ref{Renor-para}(b)-\ref{Renor-para}(d)].

To estimate the change in the ratio between spin-spin interactions $D/J_1$ within the current technology, let us consider the typical experimental parameters $U \approx 3$ eV, $t_1 \approx 0.5$ eV, $\hbar \omega \approx 0.85U = 2.55$ eV, and $a=4$ \AA, and consider an electric field amplitude of $E_0=10^9$ V/m. Using these parameters, the DMI and the exchange interaction are renormalized as $\tilde{D}/D=1.047$ and $\tilde{J}_1/J_1=1.016$, respectively. Thus, the ratio between these two spin-spin interactions is also renormalized as $\tilde{D}/\tilde{J}_1 = 1.031 (D/J_1)$. These values are detectable experimentally. In Ref. \cite{Mikhaylovskiy2015}, a change of $0.01\%$ in the ratio between the DMI and the exchange interaction has been reported by measuring the photoexcitation of the quasiantiferromagnetic mode in FeBO$_3$.

Equations (\ref{MKMHeffw}) and (\ref{Renor-para}) explicitly show that the spin Hamiltonian in the presence of a time-dependent external field can be effectively written as $\tilde{H}_{\text{S}}=H_{\text{S}}+g_{\alpha \beta i j} S^\alpha_i S^\beta_j E^{\alpha} E^{*\beta} $, where $\alpha$ and $\beta$ represent the spatial components of vectors, $i$ and $j$ refer to lattice sites, and $g$ is the optomagnetic coupling tensor, which can be read off from Eq. (\ref{Renor-para}). Thus, the dielectric permittivity tensor, which determines the optical properties of the medium, is given by $\varepsilon_{\alpha \beta}=\partial^2 \tilde{H}_{\text{S}}/\partial E^{\alpha}\partial E^{*\beta}$. The optomagnetic effect, which is described by the dielectric permittivity $\varepsilon$, can be detected by measuring the intensity of the light scattered by magnons, $I_{\mathrm{sc}} \propto (\varepsilon_{\alpha \beta} E_0)^2$ \cite{Demokritov1985}.

In ultrafast spin dynamics experiments, very intense laser pulses are used, and thus, it might be relevant to consider how heating might affect the validity of our approach. Recent theoretical \cite{2016AnPhy.367...96K,Rudner2019,PhysRevB.97.014311} and experimental \cite{PhysRevA.96.053602} works have shown that the energy absorption rate is exponentially suppressed for high-frequency laser pulses, i.e., for $\hbar \omega/W\gg 1$, where $W\propto t_1$ is the fermionic bandwidth, and this condition holds in ultrafast experiments with optical laser pulses. Thus, rapidly driven systems have a very long prethermalization period, implying that the evolution of these systems in the presence of short laser pulses can be safely described by our formalism.

In summary, we have investigated the effect of intense high-frequency polarized laser pulses on 2D AFM spin-orbit Mott insulators using the Floquet theory. We have found that both the sign and the amplitude of the ratio between the DMI and the exchange interaction in a honeycomb lattice can be modified, regardless of the helicity of the laser pulse. In general, we have shown that this ratio might be renormalized only in systems with special lattice symmetries in which the DMI and Heisenberg exchange interaction in the spin Hamiltonian are originated from hopping integrals between different distance sites in the electronic Hamiltonian.
Our calculations propose another way to achieve the ultrafast and energy-efficient control of spin-spin interactions and thus the engineering of topological objects and the topological properties of 2D van der Waals magnetic materials. The possibility of the ultrafast optical modification of the exchange interaction in bulk iron oxides has recently been reported \cite{Mikhaylovskiy2015}. We hope that our work will motivate different ultrafast experiments on measuring both the exchange interaction and the DMI in 2D magnetic systems.

\textit{Acknowledgments.} The research leading to these results was supported by the European Research Council via Advanced Grant No. 669442, ``Insulatronics,'' and by the Research Council of Norway through its Centres of Excellence funding scheme, Project No. 262633, ``QuSpin.''

\bibliographystyle{apsrev4-1}
\bibliography{paper}

\end{document}